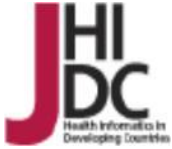

# Technological Utilization in Remote Healthcare: Factors Influencing Healthcare Professionals' Adoption and Use

**Avnish Singh Jat[1,*], Tor-Morten Grønli [1], George Ghinea[2]**

[1] *School of Economics, Innovation, and Technology, Kristiania University College, Oslo, Norway.*

[2]*Department of Computer Science, Brunel University, Uxbridge, UK.*

**Abstract**

**Background**: This study focuses on evaluating the perceptions, competencies, and acceptance of remote healthcare monitoring technology among healthcare professionals in developing countries, specifically India and Bhutan, with an emphasis on sustainability as a crucial factor. The study aims to understand the impact of perceived ease of use, usefulness, and sustainability on the adoption of such technologies.

**Methods**: This research utilized the TAM framework to analyze the responses of 44 healthcare professionals, primarily doctors, from developing countries. The study examined participants' attitudes towards the ease of use and effectiveness of remote healthcare technologies, their intentions to use such technologies, and their perceptions of sustainable design in these technologies.

**Results**: The findings indicate a generally favorable stance towards the utility and ease of use of remote monitoring technologies in telemedicine among healthcare professionals. Participants showed positive intentions toward adopting remote healthcare technologies and acknowledged the role of technology in enhancing remote monitoring. The preference for sustainably designed technologies suggests that sustainability principles can positively influence technology acceptance. However, concerns regarding security, privacy, and network coverage emerged as significant barriers to technology adoption.

**Conclusion**: The study underscores the potential of remote healthcare monitoring technologies to improve healthcare access, especially in remote and underserved areas of developing countries. It highlights the importance of addressing security, privacy, and network infrastructure issues to enhance technology adoption. Furthermore, incorporating sustainability into the TAM provides a broader understanding of the factors influencing the acceptance and use of remote healthcare technologies among healthcare professionals.

[*] Avnish Singh Jat, School of Economics, Innovation, and Technology, Kristiania University College, Kirkegata 24, 0107 Oslo, Norway; Tel: +4741352644; E-mail: avnishsingh.jat@kristiania.no  (https://orcid.org/0000-0001-7473-3010) .

# 1. Introduction

E-visits, or electronic visits, are interactions between a patient and a provider that don't take place in person and are mostly conducted from the patient's home. Reaching patients at home is possible with remote patient monitoring (RPM) [1]. To communicate with patients and residents in their homes, physicians and public health organizations can employ mobile health. The term "mHealth" refers to the use of mobile devices to distribute health-related information. This information might be general education, targeted notifications, or communication through a health application [2]. The use of mHealth for remote monitoring allows for the collection of a patient's personal health and medical information while they are at home. Mobile devices, applications, tablets, and other smart gadgets are becoming a crucial component of mHealth, according to Becker's Health IT & CIO Report article The Rise of mHealth: 10 Trends [3]. Telecare technology includes fall detection equipment, wearable sensors, fitness trackers and apps, and medication reminders. In the 2020 issue brief from the Milbank Memorial Fund, it is discussed how treatment gaps for people with substance use disorders affect rural, minority, and vulnerable groups as well as the evidence that telemedicine is a practical means of boosting access to mental and behavioral healthcare [4]. The Tele-behavioral Health Center of Excellence, run by the Mid-Atlantic Telehealth Resource Center (MATRC), offers resources to help physicians launch or improve tele-behavioral or tele-mental health programs. The implementation of tele-behavioral health is covered in this best practice guide from the Office for the Advancement of Telehealth [5].

This article is focused on understanding the factors that influence the adoption and usage of remote healthcare technology by healthcare professionals. Through this research, we intend to examine how the perceived ease of use and usefulness of the technology impact adoption and usage and investigate how attitudes towards the technology affect adoption and usage. Research is also concerned with how designing remote healthcare technology in a more sustainable way can increase its acceptance. By exploring these different factors, we want to gain insights into how to improve the design and implementation of remote healthcare technology and increase its adoption by healthcare professionals. Exploring these factors is important because remote healthcare technology has the potential to transform the way healthcare is delivered, making it more accessible, affordable, and efficient. However, for this potential to be realized, healthcare professionals need to adopt and use the technology. Understanding the factors that influence their adoption and usage is crucial for the successful implementation of remote healthcare technology. By answering these questions, we can identify the barriers to adoption and design solutions that address them. This can lead to more effective and sustainable use of remote healthcare technology, improving healthcare outcomes and increasing access to care for patients.

Through this article, we also tried to explore the increasing importance of the role of software sustainable development which refers to the practice of incorporating sustainability principles into the software development process. It involves considering economic, social, environmental, and technical aspects of sustainability in software development [6] [7]. The awareness and adoption of sustainability concepts in software development among practitioners is still limited [8] [9]. However, there is a growing understanding that sustainability should be treated as a quality attribute and integrated into the software development life cycle [10]. The acceptance of new technologies is influenced by factors such as acceptance and sustainability. To assess the future of technology, an Integrated Acceptance and Sustainability Assessment Model (IASAM) has been developed, which allows for interactive and real-time assessment and forecasting of technology sustainability [6]. In the context of Open Science, software is recognized as an important means and method for data-driven science, and efforts are being made to integrate software into a sustainable research infrastructure.

In our research, we delved into various technology acceptance models to pinpoint the factors that influence healthcare professionals' perceptions regarding ease of use and the overall usefulness of technology. While several models were explored which are mentioned in the next section, for the purposes of this study, we primarily employed the Technology Acceptance Model (TAM). There are several reasons why conducting research on the Technology Acceptance Model (TAM) in the context of remote healthcare is beneficial. Remote healthcare technology is becoming increasingly prevalent to improve access to healthcare, particularly in rural or underserved areas. Understanding how patients and healthcare providers interact with and perceive this technology can inform its design and implementation, leading to better adoption and more effective use [11]. Our utilization of TAM aims to unravel the determinants impacting healthcare professionals' acceptance of remote healthcare technologies, with an acute focus on their perceptions of usability and utility. TAM's application in this context aims to enhance technology design and implementation for better adoption, particularly in underserved areas, thereby improving patient outcomes [12]. Furthermore, our literature review, detailed in the third section, emphasizes the model's relevance in assessing healthcare professionals' attitudes towards this technology and explores how integrating sustainable development principles can further encourage its adoption.

## 2. Subjects and Methods

The Technology Acceptance Model (TAM) was first proposed by Fred Davis and Richard Bagozzi in 1989 [13]. Since then, many researchers have developed and enhanced the model to better fit different technological contexts, and to include additional factors that influence technology acceptance. One of the most significant enhancements to TAM is the inclusion of the construct of perceived behavioral control, which represents the degree to which an individual believes they have control over using the technology. This construct was first proposed by Ajzen and Madden (1986) and later included in an extended version of TAM known as the Theory of Planned Behavior (TPB) by Ajzen (1991) [14]. Another enhancement to TAM is the inclusion of the construct of social influence, which represents the degree to which an individual is influenced by others in their decision to use technology. This construct was first proposed by Venkatesh and Davis (2000) in their extension of TAM known as the Unified Theory of Acceptance and Use of Technology (UTAUT) [15]. Additionally, several researchers have applied and adapted TAM to specific technological contexts such as mobile devices, e-commerce, and healthcare. For example, a study by G. Shin et al. (2019) applied TAM to investigate the acceptance and use of a remote monitoring system for patients with heart failure [16]. The TAM model posits that two key factors, perceived usefulness, and perceived ease of use, influence an individual's intention to use a technology, which in turn influences the actual use of the technology. Perceived usefulness is the degree to which an individual believes that using the technology will enhance their job performance or task accomplishment. Perceived ease of use is the degree to which an individual believes that using the technology will be free of effort [15]. The unit of analysis in this research is the healthcare professionals who are the potential users of remote healthcare technology. Research focuses on understanding how health professionals form attitudes towards and make decisions about using new advancements in remote healthcare technology. So, the unit of analysis in TAM is the individual user, and the model focuses on understanding how individuals form attitudes toward and make decisions about using new technology.

**2.1 Research Questions**

The research aims to answer the following research question which was elaborated in the study:

*2.1.1. How do healthcare professionals' perceptions of the ease of use and usefulness of remote healthcare technology influence their adoption and usage of the technology?*

*2.1.2. How do healthcare professionals' attitudes towards remote healthcare technology influence their adoption and usage of the technology?*

*2.1.3. If remote healthcare technology designed in a more sustainable way increase its acceptance?*

By answering these questions, we want to identify the factors that are most important in promoting the acceptance and usage of remote healthcare technology and develop strategies to increase its adoption and effectiveness. We did a Likert scale analysis based on the Technology Acceptance Model; the following questions were asked from health professionals on 7-point Likert scale:

Table (1) Research Survey Questionnaire

| | | |
|---|---|---|
| Q1. | I overall trust the usefulness of remote monitoring devices in telemedicine. | Perceived Usefulness |
| Q2. | The use of remote monitoring device is the only means to share the body vital information in remote areas. | Perceived Usefulness |
| Q3. | I think using smart devices for remote healthcare monitoring will improve the performance of remote healthcare. | Perceived Usefulness |
| Q4. | I think utilizing remote monitoring devices in remote healthcare assistance is useful. | Perceived Usefulness |
| Q5. | I think utilizing remote monitoring devices in remote healthcare assistance has some advantages. | Perceived Usefulness |
| Q6. | I think utilizing remote monitoring devices will simplify the process of remote health consultation. | Perceived Ease of Use |
| Q7. | I think the process of providing healthcare assistance virtually is easy. | Perceived Ease of Use |
| Q8. | It would be easy for me to learn to obtain healthcare data through smart dices for remote healthcare. | Perceived Ease of Use |
| Q9. | My interaction with a mobile phone to provide remote healthcare assistance would be clear and understandable. | Perceived Ease of Use |
| Q10. | I intend to use remote healthcare assistance that can enable to provide patient vitals. | Behavioral Intention to Use |
| Q11. | I predict that I should use vital signs remotely monitored while giving remote healthcare assistance. | Behavioral Intention to Use |
| Q12. | I think remote healthcare assistance through smart devices is secure. | Security and Privacy(E.V.) |
| Q13. | I think utilizing mobile phone for remote healthcare monitoring does not disclose my private information. | Security and Privacy(E.V.) |
| Q14. | People who influence my behavior think that I can use the mobile phone and smart devices to provide remote healthcare assistance. | Social Influence(E.V.) |
| Q15. | People who are important to me think that I can use the mobile phone and smart devices to provide remote healthcare assistance. | Social Influence(E.V.) |
| Q16. | Network coverage is good in remote areas to provide the remote healthcare. | Technology Support(E.V.) |
| Q17. | Technologies advances in Internet security, IoT, Machine Learning will facilitate more effective remote health care monitoring. | Technology Support(E.V.) |
| Q18. | I prefer to use the software designed on sustainable software engineering principles. | Sustainable Development Impact(E.V.) |
| Q19. | I would more likely to use remote healthcare assistance software which is as cost-efficient, productive, and eco-friendly. | Sustainable Development Impact(E.V.) |

**2.2 Sampling**

Researchers have considered the strengths and limitations of quota sampling and use it in conjunction with random sampling, to ensure that the sample is representative of the population. Quota sampling is a method of sampling in which the researcher sets a target or quota for certain characteristics of the sample, in this research their profession and then we recruited participants to meet the quotas. Quota sampling is a non-probabilistic sampling method where the researcher specifies a certain number or proportion of participants to be selected from specific subgroups within the population. In alignment with the focus of our research, we specifically targeted health professionals who are currently engaged in, or have the potential to engage in, Remote Healthcare services. To provide further specificity, the primary geographical locations of these doctors were from two prominent hospitals in India: Shri Mahant Indiresh Hospital in Dehradun and Sawai Man Singh Hospital (SMS) in Jaipur. Additionally, we also reached out to their professional networks spanning other regions within India and in Bhutan, tapping into a broader spectrum of experienced individuals within the field of remote healthcare. This allowed us to capture diverse insights and nuances related to our study's context. One of the main advantages of quota sampling is that it allows the researcher to ensure that the sample is representative of the population in terms of certain characteristics [17][18].

**2.3  Data Collection Method**

We used Survey as a data collection method, Surveys are one of the most common data collection methods used in research that utilizes TAM. Surveys are a useful tool for collecting data on participants' perceptions of the technology, such as perceived usefulness and perceived ease of use [19]. Our survey was hosted on the online platform Nettskjema, employing closed-ended questions formatted on a Likert scale. The survey was available for a period of one month. To reach potential participants, we leveraged social messaging platforms, particularly WhatsApp, as our primary method of invitation.

It was imperative for us to maintain the integrity of our results; thus, we set specific qualification criteria for respondents. Only those who were at the minimum level of a resident doctor and had firsthand experience delivering healthcare services in remote regions were deemed suitable to answer the questionnaire. Any responses received from individuals who did not meet this criterion were consequently excluded. After applying these filters, we were left with feedback from forty-four qualified participants, which formed the basis of our analysis.

**2.4 Related Work**

The 2012 National Academies' workshop emphasized telehealth's potential to expand patient access, enhance care quality, and reduce costs in rural areas by preventing unnecessary ER visits and readmissions [20]. Telehealth allows rural hospitals to provide specialized care locally, thereby improving their sustainability [20]. A 2020 Journal of Rural Health article highlighted telehealth's growing utility and financial viability for providers, despite challenges like limited broadband access and technological literacy among patients [21]. Studies leveraging the Technology Acceptance Model (TAM) have shown that factors such as perceived usefulness, ease of use, enjoyment, trust, and perceived risks significantly impact the adoption of telehealth, along with demographic and socio-economic factors [22]. Research in various journals has consistently identified these factors as critical in user acceptance of remote monitoring technologies for conditions like heart failure and COPD, particularly among older adults [23-30].

Sustainable development principles in remote healthcare technology design, emphasizing reliability, efficiency, and cost-effectiveness, have been shown to enhance healthcare professionals' acceptance [41][42]. Sustainable software development practices, which focus on environmental friendliness and social responsibility, contribute to this acceptance by improving system reliability and efficiency [43][44]. The use of sustainability in healthcare technology, such as wearable sensors and eco-friendly textiles, supports broader sustainability goals and can motivate healthcare professionals through a commitment to social responsibility [45-48].

Table (2) Reviewed Article on the Importance of Sustainable Development in Technology Acceptance

| Year | Title | Results |
|---|---|---|
| 2015 | "Organizational green IT adoption: concept and evidence" [31] | This study found that environmental attitudes and norms have a positive effect on the acceptance of sustainable technologies. |
| 2015 | "Theoretical model for green information technology adoption" [32] | The study aims to understand the factors that influence the adoption of green information technology, such as environmental concerns, organizational benefits, and individual benefits. The paper provides insights into how organizations can effectively adopt green information technology to enhance their environmental sustainability and achieve long-term benefits. |
| 2018 | " Smart quintuple helix innovation systems: How social ecology and environmental protection are driving innovation, sustainable development and economic growth." [33] | This study found that social norms and self-efficacy (the belief in one's ability to perform a task) have a positive effect on the adoption of sustainable practices. |
| 2015 | " Psychological Factors Influencing the Managers' Intention to Adopt Green IS: Concepts, Methodologies, Tools, and Applications" [34] | This study found that perceived behavioral control, or the belief that one has the ability to perform a behavior, plays a significant role in the adoption of green IT (information technology) practices. |
| 2022 | "Environmental sustainability: A technology acceptance perspective" [35] | This study examines the role of technology in promoting environmental sustainability from the perspective of technology acceptance. It explore how individuals and organizations perceive and adopt technologies that support environmental sustainability, and how factors such as perceived usefulness, perceived ease of use, and attitudes towards technology influence technology adoption for sustainability purposes. |
| 2019 | "Introducing a three-tier sustainability framework to examine bike-sharing system use: An extension of the technology acceptance model" [36] | The authors extend the Technology Acceptance Model (TAM) to include three tiers: environmental, social, and economic sustainability. The aim of this study is to better understand the factors that influence the adoption and continued use of bike-sharing systems, considering the impact of these systems on the environment, society, and the economy. |
| 2017 | "Perceived ease of use and usefulness of sustainability labels on apparel products: application | The study explores consumers' perceptions and attitudes towards sustainability labels, and how these influence their buying behavior. The paper applies the Technology Acceptance Model (TAM) to examine the |

| | of the technology acceptance model" [37] | relationship between perceived ease of use and perceived usefulness of sustainability labels. |
|---|---|---|
| 2022 | "Searching for New Technology Acceptance Model under Social Context: Analyzing the Determinants of Acceptance of Intelligent Information Technology in Digital Transformation and Implications for the Requisites of Digital Sustainability" [38] | The study aims to identify the factors that contribute to the acceptance of intelligent information technology, such as social influence, perceived ease of use, and perceived usefulness, and to determine the implications of these factors for digital sustainability. The paper provides insights into how organizations can effectively adopt intelligent information technology to enhance their digital sustainability. |

The results of the studies listed above generally show that sustainable development has a positive impact on technology acceptance. The findings suggest that factors such as social influence, perceived benefits, and perceived ease of use play important roles in the adoption of green and renewable energy technologies. The studies also suggest that sustainable development initiatives can increase awareness and understanding of the benefits of these technologies, leading to greater adoption and acceptance. However, some of the studies highlight the importance of addressing potential barriers, such as cost and lack of infrastructure, to achieve widespread technology acceptance and adoption.

## 3. Results

In this study, the Technology Acceptance Model (TAM) by Venkatesh and Davis (1996) serves as the foundation for understanding the factors that influence the adoption of a new technology. The TAM model posits that Perceived Usefulness and Perceived Ease of Use are the two primary determinants of an individual's Intention to Use a technology. In this context, the study has four latent variables: External Variables (8 items), Perceived Usefulness (5 items), Perceived Ease of Use (4 items), and Intention to Use (2 items). The R-square values obtained for Perceived Usefulness (0.437), Perceived Ease of Use (0.347), and Intention to Use (0.405) indicate the proportion of variance in these latent variables that can be explained by their respective predictors.

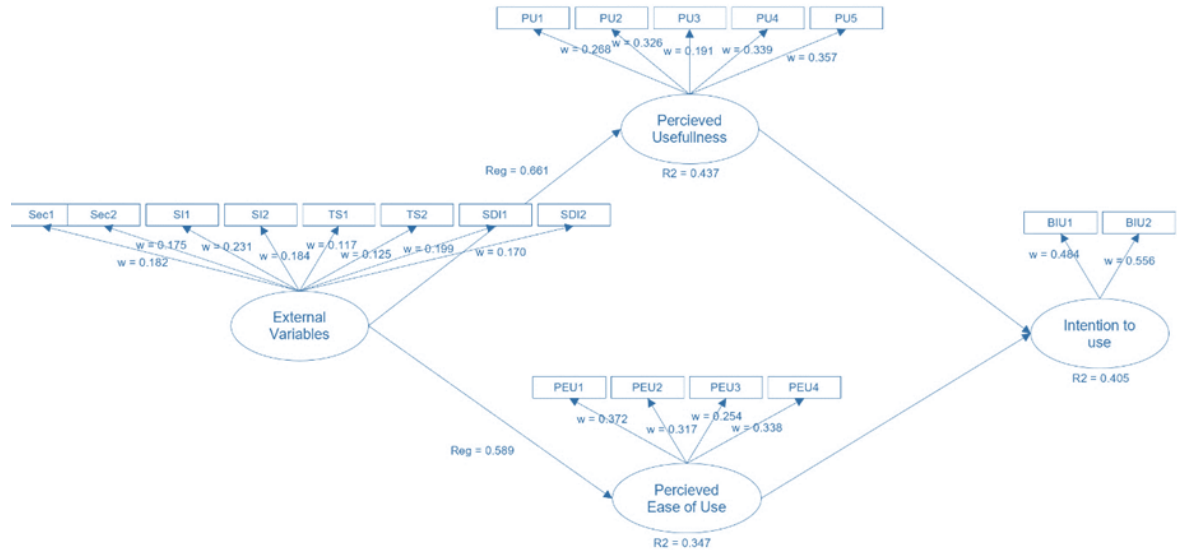

Figure (1) Technology Acceptance Model based on TAM model by Venkatesh and Davis, 1996.

**Referring to the TAM model, the results of the study suggest the following**:

a) External Variables have a significant influence on both Perceived Usefulness (regression = 0.661) and Perceived Ease of Use (regression = 0.589). This indicates that external factors play a crucial role in shaping the perception of a technology's usefulness and ease of use. The study extends the TAM model by incorporating the impact of External Variables, highlighting their importance in understanding technology adoption.

b) The R-square value of 0.437 for Perceived Usefulness implies that 43.7% of the variation in this variable can be explained by the model, primarily driven by the influence of External Variables. This suggests that the model is moderately effective in explaining Perceived Usefulness, but other factors may also contribute to the variance.

c) The R-square value of 0.347 for Perceived Ease of Use indicates that 34.7% of the variation in this variable can be accounted for by the model, with External Variables playing a significant role. Similar to Perceived Usefulness, this suggests that the model is moderately effective in explaining Perceived Ease of Use, and other factors may also be involved.

d) The R-square value of 0.405 for Intention to Use shows that 40.5% of the variance in this variable can be explained by the model, which includes the influences of Perceived Usefulness and Perceived Ease of Use. This implies that the model is moderately effective in explaining Intention to Use, and other factors not captured in the TAM model might also contribute to the variance.

For testing reliability, Cronbach's alpha test was done Cronbach's alpha is the most common measure of internal consistency ("reliability"). The result is mentioned in Table 3.

Cronbach's Alpha and Cronbach's Alpha Based on Standardized Items are measures of internal consistency or reliability of a multi-item scale/questionnaire [39]. the reliability statistics results of the study indicate that the questionnaire items are generally reliable in measuring the four latent variables. The high Cronbach's Alpha values for External Variables, Perceived Ease of Use, and Intention to Use suggest strong internal consistency, while Perceived Usefulness exhibits acceptable consistency. These results support the reliability of the questionnaire and the validity of the underlying constructs in the study.

Table (3) Reliability Test

| Latent Variables | Cronbach's Alpha | N of Items |
|---|---|---|
| External Variables | 0.9 | 8 |
| Perceived Usefulness | 0.7 | 5 |
| Perceived Ease of Use | 0.8 | 4 |
| Intention to use | 0.9 | 2 |

The cross-loadings table presents the correlations between the manifest variables (questionnaire items) and the four latent variables: External Variables, Perceived Usefulness, Perceived Ease of Use, and Intention to Use. The purpose of examining cross-loadings is to assess the convergent and discriminant validity of the items and latent variables in the study.

Table (4) Cross Loadings

| | Cross-loadings (Monofactorial manifest variables): | | | |
|---|---|---|---|---|
| | External Variables | Perceived Usefulness | Perceived Ease of Use | Intention to use |
| Sec1 | **0.672** | 0.353 | 0.562 | 0.474 |
| Sec2 | **0.723** | 0.360 | 0.518 | 0.499 |
| SI1 | **0.818** | 0.599 | 0.564 | 0.661 |
| SI2 | **0.681** | 0.577 | 0.347 | 0.475 |
| TS1 | **0.421** | 0.287 | 0.301 | 0.235 |
| TS2 | **0.678** | 0.381 | 0.246 | 0.672 |
| SDI1 | **0.825** | 0.654 | 0.346 | 0.633 |
| SDI2 | **0.819** | 0.461 | 0.396 | 0.647 |
| PU1 | 0.444 | **0.717** | 0.167 | 0.264 |
| PU2 | 0.500 | **0.728** | 0.540 | 0.359 |
| PU3 | 0.269 | **0.415** | 0.404 | 0.234 |
| PU4 | 0.444 | **0.610** | 0.388 | 0.449 |
| PU5 | 0.498 | **0.798** | 0.459 | 0.443 |
| PEU1 | 0.529 | 0.492 | **0.855** | 0.523 |
| PEU2 | 0.437 | 0.455 | **0.795** | 0.457 |
| PEU3 | 0.330 | 0.255 | **0.649** | 0.387 |
| PEU4 | 0.502 | 0.595 | **0.785** | 0.451 |
| BIU1 | 0.754 | 0.518 | 0.490 | **0.955** |
| BIU2 | 0.711 | 0.524 | 0.635 | **0.966** |

**A brief overview of the cross-loadings table reveals the following observations**:

Most items have higher loadings on their corresponding latent variables, indicating convergent validity. For instance, items PU1, PU2, and PU5 have higher loadings on Perceived Usefulness, while PEU1, PEU2, and PEU4 have higher loadings on Perceived Ease of Use.

Some items show moderately high cross-loadings with other latent variables. For example, SI1 and SI2 have high loadings on External Variables, but they also exhibit moderate loadings on Perceived Usefulness and Intention to Use. This might suggest that these items are not as discriminant as desired, and their wording or context may need further refinement.

A few items, such as TS1 and PU3, have relatively low loadings on all latent variables. This could imply that these items are not adequately capturing the intended underlying constructs or that their formulation might be unclear or ambiguous to respondents.

In summary, the cross-loadings table suggests that most manifest variables are generally converging on their corresponding latent variables, demonstrating convergent validity. However, some items exhibit moderate cross-loadings with other latent variables, indicating that there might be room for improvement in terms of discriminant validity. Refining the wording or context of these items could enhance the discriminant validity and overall quality of the questionnaire.

We performed the normality test and found the data to be not normally distributed. The test of normality is mentioned in Table 4. The Shapiro-Wilk test is a statistical test used to assess normality of a data set.[40] The "Sig." (significance) value of less than 0.001 indicates that the data deviates significantly from a normal distribution. In other words, it suggests that the data is not normally distributed and using parametric tests is not appropriate.

Table (5) Importance performance matrix analysis (IPMA)

| IPMA (Latent variable: Perceived Usefulness): | | |
|---|---|---|
| Latent variable | Importance | Performance |
| External Variables | 0.661 | 63.730 |
| **IPMA (Latent variable: Perceived Ease of Use):** | | |
| Latent variable | Importance | Performance |
| External Variables | 0.589 | 63.730 |

The Importance-Performance Matrix Analysis (IPMA) for both Perceived Usefulness and Perceived Ease of Use provides valuable insights into the importance and performance of External Variables in influencing these two outcomes (table 5).

External Variables have a moderate to strong impact on both Perceived Usefulness (Importance = 0.661) and Perceived Ease of Use (Importance = 0.589). This suggests that addressing External Variables can significantly affect the perception of the technology's usefulness and ease of use. However, the Performance score of External Variables (63.730) indicates that there is room for improvement in managing these variables to enhance both Perceived Usefulness and Perceived Ease of Use.

Table (6) Descriptive Analysis

|  | Mean | Std. Deviation | Strongly Agree | Agree | Somewhat Agree | Neutral | Somewhat Disagree | Disagree | Strongly Disagree |
|---|---|---|---|---|---|---|---|---|---|
| 1 | 6.18 | 0.691 | 14 | 25 | 4 | 1 | 0 | 0 | 0 |
| 2 | 5.93 | 1.087 | 17 | 13 | 9 | 4 | 1 | 0 | 0 |
| 3 | 6.30 | 0.632 | 17 | 23 | 4 | 0 | 0 | 0 | 0 |
| 4 | 6.18 | 0.657 | 14 | 24 | 6 | 0 | 0 | 0 | 0 |
| 5 | 6.32 | 0.771 | 22 | 14 | 8 | 0 | 0 | 0 | 0 |
| 6 | 6.64 | 0.574 | 30 | 12 | 2 | 0 | 0 | 0 | 0 |
| 7 | 5.23 | 1.054 | 3 | 17 | 15 | 5 | 4 | 0 | 0 |
| 8 | 5.59 | 1.019 | 6 | 23 | 8 | 5 | 2 | 0 | 0 |
| 9 | 6.07 | 1.065 | 20 | 12 | 8 | 3 | 1 | 0 | 0 |
| 10 | 6.00 | 0.915 | 14 | 19 | 9 | 1 | 1 | 0 | 0 |
| 11 | 5.75 | 1.081 | 12 | 16 | 11 | 3 | 2 | 0 | 0 |
| 12 | 4.95 | 1.140 | 2 | 14 | 15 | 6 | 7 | 0 | 0 |
| 13 | 5.16 | 1.256 | 4 | 18 | 10 | 6 | 5 | 0 | 0 |
| 14 | 6.09 | 0.936 | 17 | 18 | 5 | 4 | 0 | 0 | 0 |
| 15 | 6.16 | 0.914 | 18 | 19 | 3 | 4 | 0 | 0 | 0 |
| 16 | 4.75 | 0.991 | 0 | 12 | 14 | 13 | 5 | 0 | 0 |
| 17 | 6.48 | 0.762 | 27 | 12 | 4 | 1 | 0 | 0 | 0 |
| 18 | 6.18 | 0.947 | 21 | 13 | 5 | 5 | 0 | 0 | 0 |
| 19 | 6.27 | 0.924 | 23 | 12 | 5 | 4 | 0 | 0 | 0 |

The results of this research provide insight into the participants' attitudes, perceptions, and behaviors regarding the use of remote monitoring devices for telemedicine. The mean scores and standard deviations were calculated for each of the questions and the results were analyzed in terms of the Technology Acceptance Model (TAM). Overall, the participants trust the usefulness of remote monitoring devices in telemedicine, with a mean score of 6.18 and a standard deviation of 0.691. This indicates that they believe that the use of remote monitoring devices can benefit telemedicine. Participants generally view remote monitoring devices positively, with mean scores between 6.18 and 6.64, indicating a belief in their benefits for remote health consultations. However, they find the ease-of-use moderate (mean scores 5.23 to 6.07), suggesting some difficulty in virtual healthcare assistance, though interaction with mobile phones is clearer. Their behavioral intention towards remote healthcare assistance is moderate (mean scores 6.00, 5.75), with some reservations. Security and privacy concerns are moderate (mean scores 4.95, 5.16), reflecting worries about data security and privacy in smart devices. Social influence is viewed positively (mean scores 6.09, 6.16), indicating an influence from peers to use technology for healthcare. Technology support receives moderate scores (4.75, 6.48), with concerns about network coverage but optimism about technological advancements aiding remote healthcare. Lastly, sustainable development impact is viewed positively (mean scores 6.18, 6.27), showing a preference for cost-efficient, productive, and eco-friendly software.

## 4. Discussion

The main purpose of conducting research using the Technology Acceptance Model (TAM) for remote healthcare is to understand and predict the factors that influence the acceptance and usage of remote healthcare technology by patients and healthcare professionals. By understanding these factors, researchers can develop strategies to increase the adoption and effectiveness of remote healthcare technology, which can ultimately improve healthcare outcomes and reduce costs.

The mean scores obtained for the Likert scale analysis range from 4.75 to 6.64, with a majority of the scores falling in the 6-point range (6.00 to 6.64). This suggests that overall, the participants have positive views towards the usefulness of remote monitoring devices in telemedicine, with a slightly higher agreement towards the positive aspects. In a few questions some participants have neutral or slightly negative views, as the mean scores are not at the highest end of the scale in some questions. To get a more detailed interpretation, we mentioned the distribution of the scores in Table 6.

The standard deviation of the Likert scale analysis ranges from 0.574 to 1.256, with most of the values being in the 0.9 to 1.1 range. A lower standard deviation indicates that the data points are closer to the mean, while a higher standard deviation indicates greater variability in the data. In this case, the standard deviation values suggest that there is moderate to high variability in the answers provided by the participants, meaning that there is a range of opinions on the usefulness of remote monitoring devices in telemedicine.

### 5.1. Perceived Usefulness

The latent variable "Perceived Usefulness" provides a comprehensive insight into the participants' beliefs regarding the advantages and overall effectiveness of using remote monitoring devices in telemedicine. This construct was assessed using five distinct items (Q1-Q5) in the survey. The results obtained from the survey participants indicate a predominantly positive perception concerning the usefulness of remote monitoring devices within the realm of telemedicine. A detailed analysis of the results reveals that the highest mean score of 6.64 was observed for item Q5, which demonstrates a strong agreement among the participants about the numerous benefits and advantages associated with the deployment of remote monitoring devices in remote healthcare assistance. On the other hand, the lowest mean score of 5.93 was observed for item Q2. This result signifies that although the participants consider remote monitoring devices to be useful, they do not necessarily regard them as the sole means of sharing vital information in remote areas.

A closer examination of the standard deviations for all five items, which range from 0.574 to 1.087, suggests that the responses provided by the participants display a general consistency. This observation indicates a relatively low degree of variation in their perception of the usefulness of remote monitoring devices.

These findings highlight the favorable perception held by the study population regarding the usefulness of remote monitoring devices in telemedicine. This positive perception could play a crucial role in encouraging the widespread adoption and utilization of these devices in remote healthcare settings. The information obtained from this study can be effectively employed by healthcare providers, policymakers, and technology developers to gain a deeper understanding of the potential challenges and driving forces behind the implementation of telemedicine solutions, with a particular focus on remote areas. By acknowledging these factors, stakeholders can develop strategies to overcome barriers and capitalize on facilitators to promote the successful integration of telemedicine in remote healthcare systems.

## 5.2. Perceived Ease of Use

The latent variable "Perceived Ease of Use" aims to evaluate the participants' beliefs regarding the simplicity and user-friendliness of remote monitoring devices in telemedicine. This construct was assessed using four distinct items (Q6-Q9) in the survey. The results obtained from the survey participants indicate varying perceptions about the ease of use of remote monitoring devices within telemedicine.

The highest mean score of 6.18 was observed for item Q6, which suggests that participants believe that utilizing remote monitoring devices can simplify the process of remote health consultation. The second-highest mean score, 6.07, was observed for item Q9, indicating that the participants find the interaction with mobile phones for remote healthcare assistance to be clear and understandable. On the other hand, the lowest mean score of 5.23 was observed for item Q7, which indicates that participants may have some reservations regarding the ease of providing healthcare assistance virtually. Similarly, item Q8 has a mean score of 5.59, which suggests that participants may perceive some challenges in learning to obtain healthcare data through smart devices for remote healthcare.

The standard deviations for all four items, ranging from 0.657 to 1.065, indicate that there is some variability in participants' perceptions of the ease of use of remote monitoring devices in telemedicine.

These findings suggest that while the study population generally perceives remote monitoring devices as relatively easy to use, some concerns remain regarding the ease of providing virtual healthcare assistance and learning to obtain healthcare data through smart devices. This information can be utilized by healthcare providers, technology developers, and policymakers to address these concerns and improve the user experience, potentially leading to increased adoption and utilization of remote monitoring devices in telemedicine.

**5.3. Behavioral Intention to Use**

The latent variable "Behavioral Intention to Use" evaluates the participants' inclination to adopt and utilize remote healthcare assistance and monitoring vital signs remotely. The construct was assessed using two items, with the mean scores of 6.00 and 5.75, respectively. These mean scores indicate that the participants exhibit a moderate to high level of behavioral intention towards using remote healthcare assistance and remote monitoring of vital signs. The standard deviations of 0.915 and 1.081 reveal some variability in the responses, but the overall trend leans towards a positive attitude towards remote healthcare services and technology.

The findings of the present study align with previous research investigating healthcare professionals' behavioral intentions toward remote healthcare technology. Several studies have mentioned a high intention to use telemedicine services. Park and Woo found that military doctors and nurses had a significantly higher intention to use telemedicine, especially among women, younger individuals, and military nurses [49]. Another study by Siriussawakul and Phongsatha found that patients undergoing surgery had a significant intention to use telemedicine, with variables such as social influence, trust, price, perceived usefulness, and perceived ease of use impacting their intention [50]. Salsabila and Sari also found that there was a significant relationship between social influence, perceived usefulness, trust in providers, and trust in the internet on the intention to use teleconsultation services during the COVID-19 pandemic [51].

The current study's moderate to high level of behavioral intention, as indicated by the mean scores of 6.00 and 5.75, suggests that the participants are inclined towards using remote healthcare services and monitoring vital signs remotely. Although there was some variability in the responses, the general trend implies a favorable attitude towards remote healthcare technology. By focusing on the factors that impact behavioral intention, such as perceived usefulness and the ease-of-use construct, it may be feasible to enhance the adoption and usage of remote healthcare technology, ultimately improving patient outcomes and increasing the efficiency and effectiveness of healthcare delivery.

**5.4. External Variables**

The latent variable "External Variables" evaluates the influence of various external factors on the participants' perceptions of the ease of use and usefulness of remote healthcare assistance through smart devices. This latent variable is divided into four sub-themes: Security and Privacy, Social Influence, Technology Support, and Sustainable Development Impact. Each sub-theme is assessed using specific survey questions to better understand how these external factors impact the participants' attitudes towards the adoption and usage of remote healthcare technology.

### 5.4.1. Security and Privacy

The findings of the study suggest that the participants had a lower level of trust and comfort with the security and privacy of remote healthcare monitoring through smart devices. These results are consistent with previous research that has investigated the security and privacy concerns of patients and healthcare professionals towards remote healthcare technology. Existing research literature highlights the challenges posed by technology constraints, including security and privacy issues, which can pose risks to the adoption of telemedicine [52]. The COVID-19 pandemic has further emphasized the need for trust, privacy concerns, and perceived usefulness in telemedicine adoption [53]. Studies conducted in Indonesia also identify privacy concerns as a factor affecting the adoption of telemedicine, although it does not show a significant impact on behavioral intention to use telemedicine [54]. Lower scores on security and privacy (4.95 and 5.16) indicate participant concerns about data safety in remote healthcare. Addressing these through better security measures and clear data handling policies could improve technology adoption and healthcare outcomes. The standard deviation values for these two questions are also higher, indicating more variability and less agreement among participants. It might suggest the need to address security and privacy concerns in the development and implementation of remote healthcare monitoring solutions.

### 5.4.2 Social Influence

The findings of the study suggest that the participants had a positive attitude towards the social influence of remote healthcare technology. These results are consistent with previous research that has investigated the social influence of remote healthcare technology on patient behavior. For example, The study by Woo and Dowding found that the opinion of other individuals important to the patient was associated with telehealth initiation [55]. Similarly, the study by Sharma et al. identified social norms as a factor that impacts patients' adoption of telehealth services [56]. Similarly, the study by Chung et al. focused on the role of social support for older adults in transitioning to telehealth [57]. The study indicates that participants generally trust and value the opinions of influential people in their lives concerning remote healthcare technology, with average scores of 6.09 and 6.16 reflecting this trust. Despite some variability in responses, the trend shows a positive attitude towards the role of social influence in encouraging the adoption of this technology. Emphasizing the positive views of family and friends could potentially boost the use of remote healthcare technology, leading to better healthcare outcomes.

### 5.4.3. Technology Support

The findings of the study suggest that the participants generally had positive attitudes towards the usefulness and ease of use of remote monitoring devices in telemedicine, with relatively high means for the questions related to these topics. The results of the study suggest that the participants had some concerns about the availability of network coverage in remote areas. These results suggest that there may be challenges in implementing remote healthcare technology in remote areas, where network coverage may be limited. Previous research has also highlighted the challenges of implementing remote healthcare technology in remote and underserved areas, where infrastructure and network coverage can be limited. The participants in the current study seemed to believe that technology support is supportive of the use of remote monitoring devices, with moderate to high means for these questions. This finding suggests that participants believed that technology support could help them overcome any challenges they may face when using remote monitoring devices. However, the standard deviations also indicated that there was some variability in responses, particularly regarding network coverage concerns, suggests differing levels of concern among participants. Improving network coverage and providing better technology support could boost the adoption of remote healthcare technology, especially in underserved areas.

### 5.4.4. Sustainable Development Impact

The findings of the study suggest that the participants had a positive attitude towards the sustainable development impact of remote healthcare technology. We were not able to identify previous research that has specifically investigated the perceptions of patients and healthcare professionals towards the sustainability of healthcare technology but we identified studies related to the acceptance of software sustainable development that are mentioned in the related work section. These results are consistent with previous research that has investigated the perceptions of users toward software development while keeping sustainable principles in mind.

The mean scores for the questions related to sustainable development impact in the current study, as reflected in the means of 6.18 and 6.27, suggest that the participants had a positive attitude towards the use of software developed on sustainable engineering principles in remote healthcare assistance and a preference for eco-friendly, cost-efficient, and productive software. These results suggest that participants perceived the importance of sustainable development in the design and use of remote healthcare technology. However, the standard deviations for both questions were relatively high, indicating a range of opinions among the participants. These varying opinions may be influenced by a range of factors, such as familiarity with sustainable engineering principles, perceptions of the importance of sustainability in healthcare, and awareness of the environmental impact of healthcare technology. By integrating principles of sustainable development into the design and implementation of remote healthcare technologies, healthcare organizations can encourage the use of environmentally friendly and cost-effective solutions. Recognizing and addressing the diverse perspectives of users on sustainability can further enhance the adoption of these technologies. This approach not only minimizes the environmental footprint of healthcare services but also contributes to more effective and efficient healthcare delivery. The results of the research indicate that the participants have a positive attitude towards the usefulness of remote monitoring devices in telemedicine and believe that it can improve the performance of remote healthcare. They also perceive the process of providing healthcare assistance virtually to be easy and intend to use remote healthcare assistance that can enable to provide patient vitals. The participants have concerns regarding the security and privacy of remote healthcare assistance through smart devices and believe that network coverage is not good in remote areas. They also value the use of software designed on sustainable software engineering principles and would be more likely to use remote healthcare assistance software that is cost-efficient, productive, and eco-friendly.

## 5. Conclusion

Since its inception by Fred Davis and Richard Bagozzi in 1989, the Technology Acceptance Model (TAM) has been refined by researchers to include aspects like perceived behavioral control and social influence, adapting it for various contexts such as mobile technology, e-commerce, and healthcare [13]. The TAM framework helps predict technology acceptance and usage, which we applied to assess healthcare professionals' readiness for remote healthcare monitoring technology. Our study, however, was limited by its focus on doctors, resulting in a small sample size of 44 participants.

Our study delved into healthcare professionals' perceptions of remote healthcare technologies, focusing on the Technology Acceptance Model (TAM) constructs like ease of use and perceived usefulness. The findings revealed a strong approval of the technology's functionality and simplicity, indicating that these positive perceptions could significantly influence its adoption and utilization. We also explored the participants' attitudes towards adopting remote healthcare solutions, alongside their trust in the technology's capability to enhance remote monitoring. The results underscored a favorable attitude and high trust in the technology's advantages, suggesting a potential increase in its adoption. Furthermore, our research assessed the impact of sustainable design on the acceptance of remote healthcare technology. Healthcare professionals demonstrated a preference for sustainably engineered software, implying that favorable perceptions and sustainable design principles could enhance technology acceptance.

However, the research also identifies security, privacy, and network coverage concerns as potential barriers to adoption. To increase the overall acceptance and usage of remote healthcare assistance, these concerns need to be addressed by improving security and privacy measures and ensuring adequate network coverage in remote areas. By tackling these challenges, remote healthcare technology can become a more viable option for healthcare professionals, ultimately leading to better access to healthcare services for patients, particularly those in remote or underserved areas.

# 6. Declarations

## 6.1 Abbreviations

| | |
|---|---|
| TAM | Technology Acceptance Model |
| MATRC | Mid-Atlantic Telehealth Resource Center |
| IASAM | Integrated Acceptance and Sustainability Assessment Model |
| RPM | Remote Patient Monitoring |
| TPB | Theory of Planned Behavior |
| mHealth | Mobile Health |
| UTAUT | Unified Theory of Acceptance and Use of Technology |
| COPD | Chronic Obstructive Pulmonary Disease |
| IPMA | Importance performance matrix analysis |
| PU | Perceived Usefulness |
| PEU | Perceived Ease of Use |
| BIU | Behavioral Intention to Use |
| TS | Technology Support |
| SEC | Security |
| SI | Social Influence |
| SD | Sustainable Development |
| COVID-19 | Corona Virus Disease of 2019 |

## 6.2 Conflict of Interest Statement

The authors have no conflict of interests to declare.

## 6.3 Funding Disclosure

This research did not receive any specific grant from funding agencies in the public, commercial, or not-for-profit sectors.

# 7. References

[1]. Feroz, A., Perveen, S. and Aftab, W., 2017. Role of mHealth applications for improving antenatal and postnatal care in low and middle income countries: a systematic review. BMC health services research, 17(1), pp.1-11.

[2]. Zhong, X., Hoonakker, P., Bain, P.A., Musa, A.J. and Li, J., 2018. The impact of e-visits on patient access to primary care. Health care management science, 21, pp.475-491.

[3]. The rise of mHealth: 10 trends (no date) Becker's Hospital Review. Available at: https://www.beckershospitalreview.com/healthcare-information-technology/the-rise-of-mhealth-10-trends.html (Accessed: February 7, 2023).

[4]. Lazur, B., Sobolik, L. and King, V., 2020. Telebehavioral health: An effective alternative to in-person care. New York: Milbank Memorial Fund. Available at: https://www. milbank. org/publications/telebehavioral-health-an-effective-alternative-to-in-personcare/. Accessed December, 6, p.2021.

[5]. (2022) Telebehavioral Health Center of Excellence. Available at: https://tbhcoe.matrc.org/ (Accessed: February 7, 2023).

[6]. Zane, Barkane., Egils, Ginters. (2011). Introduction to Technologies Acceptance and Sustainability Modelling.

[7]. Leila, Karita., Brunna, Caroline, Mourão., Luana, Almeida, Martins., Larissa, Rocha, Soares., Ivan, do, Carmo, Machado. (2021). Software industry awareness on sustainable software engineering: a Brazilian perspective. Journal of Software Engineering Research and Development, doi: 10.5753/JSERD.2021.742

[8]. Leila, Karita., Brunna, Caroline, Mourão., Ivan, do, Carmo, Machado. (2019). Software industry awareness on green and sustainable software engineering: a state-of-the-practice survey. doi: 10.1145/3350768.3350770

[9]. Timo, Borst. (2015). Sustainable Software as a Building Block for Open Science. doi: 10.3233/978-1-61499-562-3-31

[10]. PohWah, Khong., RongMing, Ren. (2007). User acceptance of information technologies for enterprise development. Journal of Computer Applications in Technology, doi: 10.1504/IJCAT.2007.013352

[11]. Binci, D., Palozzi, G. and Scafarto, F., 2022. Toward digital transformation in healthcare: a framework for remote monitoring adoption. The TQM Journal, 34(6), pp.1772-1799.

[12]. Zanaboni, P. and Wootton, R., 2012. Adoption of telemedicine: from pilot stage to routine delivery. BMC medical informatics and decision making, 12(1), pp.1-9.

[13]. Davis, F.D., Bagozzi, R.P. and Warshaw, P.R., 1989. User acceptance of computer technology: A comparison of two theoretical models. Management science, 35(8), pp.982-1003.

[14]. Ajzen, I., 1991. The theory of planned behavior. Organizational behavior and human decision processes, 50(2), pp.179-211.

[15]. Venkatesh, V., Morris, M.G., Davis, G.B. and Davis, F.D., 2003. User acceptance of information technology: Toward a unified view. MIS quarterly, pp.425-478.

[16]. Shin, G., Jarrahi, M.H., Fei, Y., Karami, A., Gafinowitz, N., Byun, A. and Lu, X., 2019. Wearable activity trackers, accuracy, adoption, acceptance and health impact: A systematic literature review. Journal of biomedical informatics, 93, p.103153.

[17]. Acharya, A.S., Prakash, A., Saxena, P. and Nigam, A., 2013. Sampling: Why and how of it. Indian Journal of Medical Specialties, 4(2), pp.330-333.

[18]. Moser, C.A., 1952. Quota sampling. Journal of the Royal Statistical Society. Series A (General), 115(3), pp.411-423.

[19]. Vogt, W.P., Gardner, D.C. and Haeffele, L.M., 2012. When to use what research design. Guilford Press.

[20]. Board on Health Care Services; Institute of Medicine. The Role of Telehealth in an Evolving Health Care Environment: Workshop Summary. Washington (DC): National Academies Press (US); 2012 Nov 20. Available from: https://www.ncbi.nlm.nih.gov/books/NBK207145/ doi: 10.17226/13466

[21]. Woodall, T., Ramage, M., LaBruyere, J.T., McLean, W. and Tak, C.R., 2021. Telemedicine services during COVID-19: Considerations for medically underserved populations. The Journal of Rural Health, 37(1), p.231.

[22]. Rouidi, M., Abd Elmajid, E., Hamdoune, A., Choujtani, K. and Chati, A., 2022. TAM-UTAUT and the acceptance of remote healthcare technologies by healthcare professionals: A systematic review. Informatics in Medicine Unlocked, p.101008.

[23]. Giger, J.T., Pope, N.D., Vogt, H.B., Gutierrez, C., Newland, L.A., Lemke, J. and Lawler, M.J., 2015. Remote patient monitoring acceptance trends among older adults residing in a frontier state. Computers in Human Behavior, 44, pp.174-182.

[24]. Cimperman, M., Brenčič, M.M., Trkman, P. and Stanonik, M.D.L., 2013. Older adults' perceptions of home telehealth services. Telemedicine and e-Health, 19(10), pp.786-790.

[25]. Vorrink, S., Huisman, C., Kort, H., Troosters, T. and Lammers, J.W., 2017. Perceptions of patients with chronic obstructive pulmonary disease and their physiotherapists regarding the use of an eHealth intervention. JMIR human factors, 4(3), p.e7196.

[26]. Conn, N.J., Schwarz, K.Q. and Borkholder, D.A., 2019. In-home cardiovascular monitoring system for heart failure: comparative study. JMIR mHealth and uHealth, 7(1), p.e12419.

[27]. Shin, G., Jarrahi, M.H., Fei, Y., Karami, A., Gafinowitz, N., Byun, A. and Lu, X., 2019. Wearable activity trackers, accuracy, adoption, acceptance and health impact: A systematic literature review. Journal of biomedical informatics, 93, p.103153.

[28]. Zhou, M., Zhao, L., Kong, N., Campy, K.S., Qu, S. and Wang, S., 2019. Factors influencing behavior intentions to telehealth by Chinese elderly: An extended TAM model. International journal of medical informatics, 126, pp.118-127.

[29]. Alshammari, K.A.F., 2018. Computer Anxiety of the Digital Technologies Acceptance in Saudi Arabia. International Journal of Heritage, Art and Multimedia, 1(3), pp.32-50.

[30]. Hsiao, C.H. and Tang, K.Y., 2015. Examining a model of mobile healthcare technology acceptance by the elderly in Taiwan. Journal of Global Information Technology Management, 18(4), pp.292-311.

[31]. Deng, Q. and Ji, S., 2015. Organizational green IT adoption: concept and evidence. Sustainability, 7(12), pp.16737-16755.

[32]. Asadi, S., Hussin, A.R.C., Dahlan, H.M. and Yadegaridehkordi, E., 2015. Theoretical model for green information technology adoption. ARPN Journal of Engineering and Applied Sciences, 10(23), pp.17720-17729.

[33]. Carayannis, E.G. and Campbell, D.F., 2018. Smart quintuple helix innovation systems: How social ecology and environmental protection are driving innovation, sustainable development and economic growth. Springer.

[34]. Esfahani, M.D., Nilashi, M., Rahman, A.A., Ghapanchi, A.H. and Zakaria, N.H., 2015. Psychological factors influencing the managers' intention to adopt green IS: A review-based comprehensive framework and ranking the factors. International Journal of Strategic Decision Sciences (IJSDS), 6(2), pp.28-56.

[35]. Papagiannidis, S. and Marikyan, D., 2022. Environmental Sustainability: A technology acceptance perspective. International Journal of Information Management, 63, p.102445.

[36]. Jamšek, S. and Culiberg, B., 2020. Introducing a three-tier sustainability framework to examine bike-sharing system use: An extension of the technology acceptance model. International Journal of Consumer Studies, 44(2), pp.140-150.

[37]. Ma, Y.J., Gam, H.J. and Banning, J., 2017. Perceived ease of use and usefulness of sustainability labels on apparel products: application of the technology acceptance model. Fashion and Textiles, 4, pp.1-20.

[38]. Park, I., Kim, D., Moon, J., Kim, S., Kang, Y. and Bae, S., 2022. Searching for new technology acceptance model under social context: analyzing the determinants of acceptance of intelligent information technology in digital transformation and implications for the requisites of digital sustainability. Sustainability, 14(1), p.579.

[39]. Tavakol, M. and Dennick, R., 2011. Making sense of Cronbach's alpha. International journal of medical education, 2, p.53.

[40]. Yap, B.W. and Sim, C.H., 2011. Comparisons of various types of normality tests. Journal of Statistical Computation and Simulation, 81(12), pp.2141-2155.

[41]. Arthur, Todd, Sedano. (2017). Sustainable Software Development: Evolving Extreme Programming.
[42]. Timo, Johann., Markus, Dick., Eva, Kern., Stefan, Naumann. (2011). Sustainable development, sustainable software, and sustainable software engineering: An integrated approach. doi: 10.1109/SHUSER.2011.6008495
[43]. Koushik, Maharatna., Silvio, Bonfiglio. (2013). Systems Design for Remote Healthcare. doi: 10.1007/978-1-4614-8842-2
[44]. M., Spilka., A., Kania. (2006). Application of the sustainable materials technology model.
[45]. Valentina, Trovato., Silvia, Sfameni., Giulia, Rando., Giuseppe, Rosace., Sebania, Libertino., Ada, Ferri., Maria, Rosaria, Plutino. (2022). A Review of Stimuli-Responsive Smart Materials for Wearable Technology in Healthcare: Retrospective, Perspective, and Prospective. Molecules, doi: 10.3390/molecules27175709
[46]. Ashutosh, Tiwari. (2014). Advanced Healthcare Materials.
[47]. R., van, Berkel. (2006). Innovation and technology for a sustainable materials future.
[48]. Leal Filho, W., Tripathi, S. K., Andrade Guerra, J. B. S. O. D., Giné-Garriga, R., Orlovic Lovren, V., & Willats, J. (2019). Using the sustainable development goals towards a better understanding of sustainability challenges. International Journal of Sustainable Development & World Ecology, 26(2), 179-190.
[49]. Soyeon, Park., Kyungmi, Woo. (2023). Military Doctors' and Nurses' Perceptions of Telemedicine and the Factors Affecting Use Intention.. Telemedicine Journal and E-health, doi: 10.1089/tmj.2022.0430
[50]. Arunotai, Siriussawakul., Thanawan, Phongsatha. (2022). The Intention to Use Telemedicine by Surgical Patients in Response to COVID-19. Siriraj Medical Journal, doi: 10.33192/smj.2022.95
[51]. Isna, Mutiara, Salsabila., Kurnia, Riska, Sari. (2022). Analysis of factors related to intention-to-use telemedicine services (teleconsultation) in jabodetabek residents during the covid-19 pandemic in 2021. Journal of Indonesian health policy and administration, doi: 10.7454/ihpa.v7i3.6090
[52]. Ankur, Chattopadhyay., Nahom, M, Beyene. (2023). A W3H2 Analysis of Security and Privacy Issues in Telemedicine: A Survey Study. doi: 10.1145/3564746.3587109
[53]. Linxiang, Zhu., Junwei, Cao. (2023). Factors Affecting Continuance Intention in Non-Face-to-Face Telemedicine Services: Trust Typology and Privacy Concern Perspectives. Healthcare, doi: 10.3390/healthcare11030374
[54]. Betty, Purwandari., Imairi, Eitiveni., Mardiana, Purwaningsih. (2023). Factors Affecting Adoption of Telemedicine for Virtual Healthcare Services in Indonesia. Journal of Information Systems Engineering and Business Intelligence, doi: 10.20473/jisebi.9.1.47-69
[55]. Kyungmi, Woo., Dawn, Dowding. (2020). Decision-making Factors Associated With Telehealth Adoption by Patients With Heart Failure at Home: A Qualitative Study. Cin-computers Informatics Nursing, doi: 10.1097/CIN.0000000000000589
[56]. Pooja, Sharma., Asmat, Ara, Shaikh., Ameet, Sao., Nitesh, Rohilla. (2022). Using Technology Acceptance Model, Analyzing the Role of Telehealth Services in the Healthcare Industry During COVID-19. Asia Pacific journal of health management, doi: 10.24083/apjhm.v17i2.1815
[57]. Grace, S., Chung., Chad, Ellimoottil., Jeffrey, S., McCullough. (2021). The Role of Social Support in Telehealth Utilization Among Older Adults in the United States During the COVID-19 Pandemic. doi: 10.1089/TMR.2021.0025